# 5W1H-based Expression for the Effective Sharing of Information in Digital Forensic Investigations

Jaehyeok Han[1], Jieon Kim[1], and Sangjin Lee[1]

[1]Institute of Cyber Security & Privacy (ICSP), Korea University
145, Anam-ro, Seongbuk-gu, Seoul, South Korea
{one01h,kijie,sangjin}@korea.ac.kr

**Abstract.** Digital forensic investigation is used in various areas related to digital devices including the cyber crime. This is an investigative process using many techniques, which have implemented as tools. The types of files covered by the digital forensic investigation are wide and varied, however, there is no way to express the results into a standardized format. The standardization are different by types of device, file system, or application. Different outputs make it time-consuming and difficult to share information and to implement integration. In addition, it could weaken cyber security. Thus, it is important to define normalization and to present data in the same format. In this paper, a 5W1H-based expression for information sharing for effective digital forensic investigation is proposed to analyze digital forensic information using six questions—what, who, where, when, why and how. Based on the 5W1H-based expression, digital information from different types of files is converted and represented in the same format of outputs. As the 5W1H is the basic writing principle, application of the 5W1H-based expression on the case studies shows that this expression enhances clarity and correctness for information sharing. Furthermore, in the case of security incidents, this expression has an advantage in being compatible with STIX.

## 1 Introduction

Digital forensics is a technique for finding clues and evidence for criminal investigations by collecting and analyzing data stored on digital devices. Recently, digital forensics has expanded its use in various fields such as security incidents, besides criminal investigations. Also, digital forensic tools such as EnCase, FTK (Forensic Tool Kit), and Autopsy have been developed. However, those forensic tools extract data to different outputs, which make difficult to integrate and share the digital forensic information. Since the tools use all different ways to express and present their data, it is inefficient and ineffective to process data integration. Thus, this paper suggests a methodology to improve data processing and facilitates the development of digital forensic tools by data standardization, which could help people read easier.

We proposes 5W1H-based expression for information sharing for effective digital forensic investigation. Further, the other parts of the paper are organized as follows. Section 2 reviews related work and explains a relationship of data, information, and intelligence in digital forensics. Methodology for 5W1H-based expression for information sharing is presented in section 3, and section 4 introduces case studies and applies them to 5W1H-based expression. Section 5 discusses the results and concludes.

## 2 Background and Related works

First, we looked the concepts of the data, information, and intelligence on the digital forensic investigation, information is produced by analyzing data from digital evidence, and intelligence is produced by utilizing information. In other words, after drawing results from digital evidence, a direction on how to investigate will be set. Data, information, and intelligence can be classified by their volume and usability. Data is typically available in huge volumes, and describes individual and unarguable facts. Information is produced when a series of data points are combined to answer a simple question. Intelligence takes this process a stage further by interrogating data and information to tell a story (e.g. a forecast) that can be used to derive decision making. Crucially, intelligence does not answers a simple question, rather it paints a picture that can be used to help people to answer much more complicated questions.

Information on its own may be of utility to the investigator, but when related to other information about the operational environment and considered in the light of past experience it gives rise to a new understanding of the information, which may be termed 'intelligence'. Ultimately, intelligence has two critical features that distinguish it from information. Intelligence allows anticipation or prediction of the certain situations and circumstances, and it helps to make decisions by illuminating the differences in available courses of action [12]. Fig. 1 describes when applying relationship of data, information, and intelligence on digital forensic investigation in Windows OS system. From the disk image dump file obtained in the data collection level, the files extracted through volume and file system analysis are inputted into data (registry, event log, etc). Contents, which are parsed results from files, would draw information.

In other words, intelligence is an integrated connection to information which could be generated to a report. This is because the conclusions in the report contain information that can explain the causal relationship or whether it is possible to investigate the content requested from the given data. While law enforcement and investigative agencies focus primarily on generating reports, digital forensic investigators focus on the systematic and effective method to accumulate knowledge at information level for the efficiency of resources such as profiles. In fact, benefits of a sully structured form is to automate the normalization and combination of data, and to fuse together disparate sources of information [6]. So, we suggests how to present and share information for effective digital forensic investigation based on 5W1H.

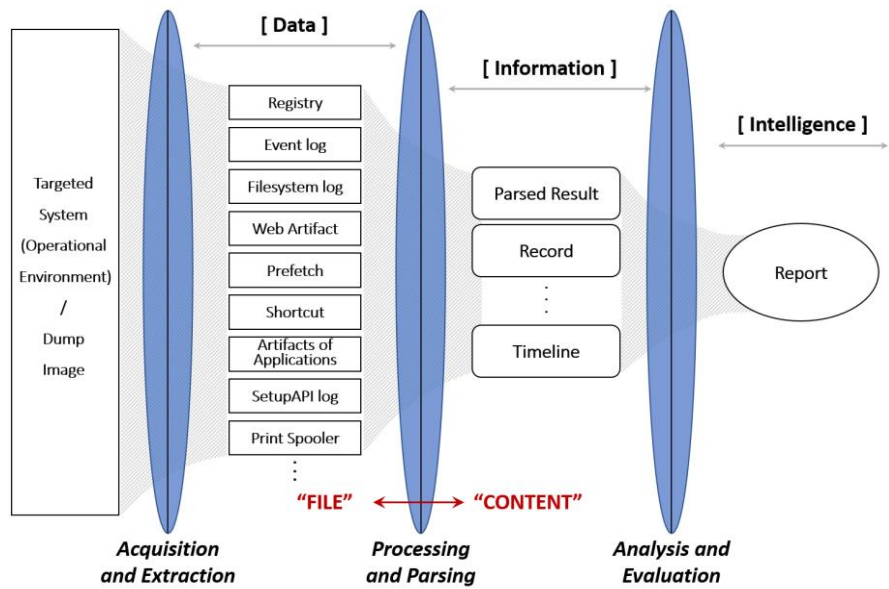

**Figure 1.** Relationship between data, information, and intelligence in a digital forensic investigation with Windows artifacts.

## 3    Information expression based on 5W1H

The primary purpose of digital evidence exchange format and package is to store and preserve the collected results from digital data. Since lots of system logs contains how the evidence is managed in detail, all the processes can be traced back if behavior and objects are guaranteed their originality. When repeating the same process over and over again, it should yield the same results. If there is a trouble while collecting and acquiring data, then the process on digital forensic investigation is meaningless after that. In addition, it would be created as a dump image or logical file, then volume is too massive to shared contents that have processed and parsed from the internals of files. So, it is necessary to downscale the range of investigation by extracting only essential contents. It is important to analyze the contents to know what information they have and the results will be different by authorities and stakeholders.

A 5W1H method is a basic writing principle, which asks six questions--what, who, where, when, why and how. Media that delivers objective facts followed this rule for correctness, conciseness, and clearness. Previous research has been studied usually in natural sentences to understand tasks and summarize the information. Information representation based on the 5W1H method in digital forensic investigation maintained their characteristics and enhanced its search efficiency and visibility. Moreover, it is a primary element for vague and complex investigation. Basically, digital forensic investigators try to list as much information as possible. For example, the header of a ZIP file contains metadata such as the name of a compressed file, modified time, and file size. That information is stored in a unit of a structure. When there is no capability to analyze it, it is unable to know what each structure points to.

The process of *parsing* is an automated system to put newly input data on an already known frame. It is a simple process to normalize it into an existing frame by statistics, empirical experiments, or logical analysis, but it can't provide any new information. In the case of information which can be expressed by 5W1H, it can ensure clarity and correctness. However, some information is unable to be expressed by 5W1H. In this case, the information only contains simple state of an object or the information cannot answer to the 5W1H questions. If no information is available to represent 5W1H-based expression, it still needs to list the normalized contents and do further examination to find traces more.

### 3.1 Information Extraction

In order to extract information from data, volume and file system analysis [3] of the targeted system (or operational environment) or dump images is processed first, and then the content is extracted through structure analysis (a.k.a.parsing) by types of files [8]. Most tools create analysis results in the form of databases, markup language, and comma-separated values (CSV). Different types of data outputs can be converted to the same type of outputs by format conversion. In the case of security incidents, STIX (Structured Threat Information eXpression) or CybOX (Cyber Observable eXpression) [2][4] can be used since it requires a faster information exchange. Besides contents extracted from files, a specification language such as UCO (Unified Cyber Ontology) [5] to express ontology can be utilized. However, many tools, used for parsing, print out results which have different keys with the same content, which makes digital forensic investigation inefficient.

For example, let's consider messages that sent and received using a messenger. Mostly messengers have log files of the activity following elements in common such as timestamp, sender, receiver, text, and attachment. These things should have examined during investigation, but the normalization of the elements is different by type of devices, operating systems, or applications. In reality, they are often stored in SQLite databases format and it is easy to meet a different schema for each types. It means that it is necessary to express the information in a standardized format such as CASE [2].

## 3.2 Classification and Filtering

Since the amount of content to be extracted from the file is very large, and the 5W1H is useful for expressing the behavior of a user or system, the information is divided into 20 categories. The category covers all the keys used in Opentext EnCase(19), X-Ways Forensics(21), Magnet AXIOM(14) [10], and Oxygen Forensics(16). Additionally, the lists necessary for the investigation is derived as follows and the 5W1H was tagged with the contents in accordance with Table 1, which is the data types and definitions for classification.

- archive
- call
- calendar
- cloud
- connection
- document
- email
- exchange
- executable
- font
- location
- media
- memory
- message
- network
- people
- social
- string
- system
- web

Also, in order to exchange information which is much larger than intelligence as a resource for decision-making, it is not possible that data is normalized in the form of collection, refinement, or package [9]. In other words, we need a virtual layer that acts as a bridge to information, and we propose a classification of information based on the 5W1H.

The information with category and the 5W1H tag is written by JSON-LD (JavaScript Object Notation for Linked Data), normalized format, and saved it with serialization for sharing or archiving with other users over a network. In the case of STIX, 94 categories are provided. It usually defined the information in network communication, and it is useful to use it in conjunction since it is very limited to express the analyzed results in digital forensic investigation. Besides, certain rules can be shared using the Indicator of Compromise (IOC). It is useful to have a way of expressing this information in order to classify and share it. Especially, the 5W1H-based method will help effective digital forensic investigation. This process often involves a large number of stakeholder personnel to ensure a chain of custody, a guarantee of participation, and credibility, which can be effective for flexible communication.

**Table 1.** Criteria for applying domains based on 5W1H.

| Domain | Concept | Question |
|---|---|---|
| who | subject | Who (or What) did it?<br>Who acted on whom? |
| when | time, period | What time did it happen?<br>Since when, until when? |
| where | place, direction | Where it was happend?<br>Where did you go from where? |
| what | object | What was it?<br>What did you get? |
| why → etc | reason | Why do you think so?<br>Where did you check it? |
| how | method | How did it work? |

### 3.3    Writing Objects for Sharing

We have defined a specification to create flexible representations of information and have correctness and conciseness on digital forensic investigation. This aims to make it effective to integrate and process information in order to derive additional clues. In our method, the basic unit of information expression is an object written in JSON (JavaScript Object Notation) format and it has attributes in key-value. A object is the normalized data to express traces or events. A object is written with contents that will come from images or files according to the type of data with documents which defines many attributes, which are materials to normalized and express contents, facilitate the exchange of information, and develop an automated system. We defines some attributes or used vocabularies already defined in STIX or DFXML(Digital Forensics XML) [11]. All objects have header to identify each of them. Depending on this attributes, they are classified two types as trace object or event object.

   A trace object describes the fact observed from images or files of analysis and is capable to represent the state of a particular subject. An event object is a model designed to represent information that can be inferred from the contents identified from the trace object. So, this supports the ability to correlate facts represented by the trace object with each other and to associate events with related objects, as shown in the Fig. 2. Here, in writing the event object, we wrote the relation to objects in the form based on 5W1H, including activities by user or system. In other words, it can be seen as a reassignment of the trace object representing the content, as shown in Fig. 3.

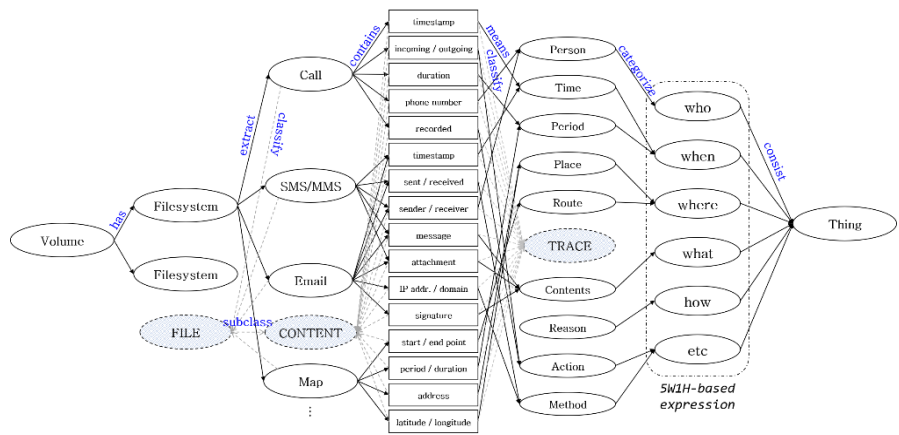

**Figure 2.** Examples of the classification of the contents for writing objects based on 5W1H categories.

- traceObj = { id: **trace1**,
              type: system,
              ArtifactType: WinEventLog,
              Even-tID: 4624,
              SubjectUserName: Jaehyeok,
              EventOccuredTime: 2019-07-14,
              source: security.evtx }
- eventObj = { id: event1,
              who: Jaehyeok,
              how: logged on,
              when: 2019-07-14,
              etc: souce = [trace1]}

⟹ An user <u>Jaehyeok</u> successfully <u>logged on</u> at <u>2019-07-14</u> by **trace1**.

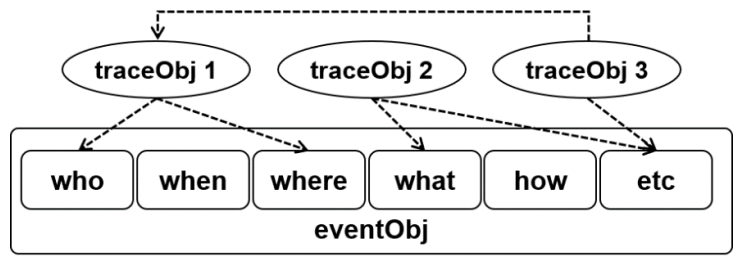

**Figure 3**. Relationship between trace object and event object to express information.

## 4 Case studies

### 4.1 Investigations into confidential leaks

In this section, we will take a look at the cases where information expression based on the 5W1H is applied. Tom was an employee in ABC company, who quitted his job after receiving a job offer from a competitor. Tom got a job offer from an employee of the competitor by email and text message before leaving the company. Tom used a USB storage device to copy *Confidential Document.docx* and *Account Book.xlsx*. He also copied the *partner evaluation report.pdf* of another department using the cloud service on the company's public PC. Tom sent a text message to the employee in the competitor who suggested a career change after leaving his office. At that time, a member of the audit team felt suspicious of Tom and investigated his devices that were given to him to determine whether he leaked the confidential information or not [13].

This is another scenario. An employee at an IT company suddenly had a health problem while working abroad. His family requested an investigation to find whether it is an industrial accident. They also requested to analyze the computer usage history within one week of the incident through digital forensics to determine whether excessive work affected his sudden death. As a result, it was found out that the user logged in to the work computer for a long time and accessed a large number of document files right before the incident.

As described above, the 5W1H-based expression is useful for applying the digital profiling technique because it specifies and stores the domain. Digital profiling [7] refers to the process of collecting, analyzing and utilizing information on cyberspace as well as digital devices. It includes information about personal information, characteristics, behaviors, affiliations, connections, and interactions and can be used in organized crime investigations, merchandise marketing, corporate security, criminal justice and human resources.

### 4.3 Verification of tools

Many tools must be used to ensure reliability [15] during the analysis process because the tools used in digital forensic investigation are different from each other and the credibility of results are not guaranteed. NIST proposed a method to verify the reliability of forensic tools. CFTT [14] provides the information necessary for the tool user to determine the reliability of the tool. There are disk imaging, write block, deleted file recovery, and forensic file carving, but the methodology for analytical tool testing from the acquired data is limited. This is because it is difficult to establish a clear standard for evaluating the reliability of the analysis tool that grasps the behavior of the system or the user's behavior

| Key | Output of Tool "A" | Offset/size | Ans. | Output of Tool "B" | Offset/size | Ans. |
|---|---|---|---|---|---|---|
| who: From | Tom | 10h / 3 | ○ | Jane | 10h / 3 | × |
| who: To | Jane | 20h / 4 | ○ | - | - | △ |
| when: StartTime | 2020-02-10 19:01:15 | 28h / 8 | ○ | 2020-02-10 19:01:15 | 28h / 8 | ○ |
| when: EndTime | 2020-02-01 19:12:27 | 30h / 8 | ○ | - | - | △ |
| how: Application | Google Hangouts | 40h / 9 | ○ | Google Hangouts | 40h / 9 | ○ |

call_log.db — Database that stores call logs on Android devices

*input* → *input* →

○ : Output is correct (True positive)
△ : No results (Not supported)
X : Output is incorrect (False positive)

**Figure 4**. Comparison of analysis results for different tools using 5W1H-based expression.

through an evaluation image. Therefore, if the analysis results extracted from different tools are compared to each other based on 5W1H method proposed in the paper, it can be used to verify the reliability of each tool.

Fig. 4 is an example how to determine the reliability of tools by comparing the analysis results of different tools. For the sake of illustration, *Tool A* and *Tool B* shows in the *call_log.db* as a call log data for who–From, To, When–StartTime, EndTime, and how–Application. Contents corresponding to CallType was extracted from *Tool A* and that result is correct–All answer is Ｏ and this means TRUE. On the other hand, *Tool B* derived several incorrect answer or nothings. So, we could consider that *Tool A* is more reliable tool than *Tool B*.

## 5    Conclusion

The types of files covered by the digital forensic investigation are wide and varied, however, there is no way to express the results into a standardized format. Different outputs make it time-consuming and difficult to share information and to implement integration. Based on the 5W1H-based expression, digital information from different types of files is converted and represented in the same format of outputs. The application of the 5W1H-based expression on the case studies shows that this expression enhances clarity and correctness for information sharing. In addition, This 5W1H based expression improved from syntactic system to semantic system.

It is best to standardize the format of data expressions and to normalize the format when outputting results, but since many tools have already been developed and distributed, it is inefficient and practically impossible to force standardization to them. We have developed a prototype mapper script using python that can convert the output from Plaso [4] to ours. This mapper is one of our efforts to express information from various contents, to comprehensively analyze, and profile them suitably. In the future, the range of data storage will expand to cyberspace including digital devices. With this trend, attentions and efforts will be needed to develop and standardize the specification language.